# Deep-learning based down-scaling of summer monsoon rainfall data over Indian region

**Bipin Kumar\*, Rajib Chattopadhyay, Manmeet Singh, Niraj Chaudhari, Karthik Kodari and Amit Barve**

**Bipin Kumar, Rajib Chattopadhyay, Manmeet Singh**
Indian Institute of Tropical Meteorology, Ministry of Earth Sciences, Dr. Homi Bhabha Road, Pashan, 411008, India

**Manmeet Singh**
IDP in Climate Studies, Indian Institute of Technology Bombay, India

**Niraj Chaudhari, Amit Barve**
Ramrao Adik Institute of Technology, Nerul, Navi Mumbai, MH, 400706, India

**Karthik Kodari**
Defence Institute of Advanced Technology, Girinagar, Pune, MH, 411025 India

**\*Corresponding Author**
**Bipin Kumar**
Indian Institute of Tropical Meteorology,
Dr. Homi Bhabha Road,
Pashan, 411008,
India
E-mail: bipink@tropmet.res.in

**Abstract**

Downscaling is necessary to generate high-resolution observation data to validate the climate model forecast or monitor rainfall at the micro-regional level operationally. Dynamical and statistical downscaling models are often used to get information at high-resolution gridded data over larger domains. As rainfall variability is dependent on the complex Spatio-temporal process leading to non-linear or chaotic Spatio-temporal variations, no single downscaling method can be considered efficient enough. In data with complex topographies, quasi-periodicities, and non-linearities, deep Learning (DL) based methods provide an efficient solution in downscaling rainfall data for regional climate forecasting and real-time rainfall observation data at high spatial resolutions. In this work, we employed three deep learning-based algorithms derived from the super-resolution convolutional neural network (SRCNN) methods, to precipitation data, in particular, IMD and TRMM data to produce 4x-times high-resolution downscaled rainfall data during the summer monsoon season. Among the three algorithms, namely SRCNN, stacked SRCNN, and DeepSD, employed here, the best spatial distribution of rainfall amplitude and minimum root-mean-square error is produced by DeepSD based downscaling. Hence, the use of the DeepSD algorithm is advocated for future use. We found that spatial discontinuity in amplitude and intensity rainfall patterns is the main obstacle in the downscaling of precipitation. Furthermore, we applied these methods for model data postprocessing, in particular, ERA5 data. Downscaled ERA5 rainfall data show a much better distribution of spatial covariance and temporal variance when compared with observation.

## 1. Introduction

Downscaling of rainfall data over Indian region during the summer monsoon season is difficult as the summer monsoon rainfall results from a multi-scale spatio-temporal dynamical process with significant variance (Chang et al. 2018). It is a complex interacting system with various climate teleconnections playing a considerable role in rainfall variabilities such as ENSO(Sikka 1980; Pant and Parthasarathy 1981), PDO (Krishnan and Sugi 2003), IOD (Ashok et al. 2001), and Volcano (Singh et al. 2020). Hence, regional variations of monsoon rainfall are often quite large varying from a few millimeters to thousands of millimeters within a few hundred kilometers (e.g., near the Western Ghats) and over several belts of the Himalayas. Figure 1 (a,b) shows the high degree of spatial dispersion and spatial gradient in the distribution along the western ghat region and high gradient in monthly averaged rainfall. Monsoon rainfall can be classified into several coherently fluctuating zones, which may be linked to complex multi-scale processes (Gadgil et al. 1993; Gadgil 2003; Moron et al. 2017).

Incoherent fluctuations, extreme flood or deficient rainfall events, and high-impact rainfall events often cause severe damage and property loss in the Indian subcontinent during the monsoon season. The Indian monsoon rainfall is mainly affected/dominated by orography in the Western Ghats region and Himalayan area of North and North-East of India

The high impact rainfall events are often localized and associated with large-scale meteorological systems, for instance, depressions and cyclones. Policymakers and disaster managers are usually interested in future outlooks of such extreme rainfall events from several monsoon season vagaries. In the context of climate change, where it is projected that extreme rainfall events are increasing during the monsoon season, the downscaling assumes a significant area of active research.

Meteorological forecasts and observations are often generated at a very coarse spatial resolution. While technology is available to perform dynamic forecasts at high spatial resolutions, it can be very compute-intensive and relatively expensive to manage the operational runs for such high spatial resolutions.

The generation of low-resolution climate forecasts (e.g., (Swapna et al. 2018), (Krishnan et al. 2019)) can be downscaled (von Storch et al. 1993; Wilby and Dawson 2013) for better regional outlook generation.

Similarly, downscaling of observed precipitation data can also be made based on a low-resolution observation (Sahai et al. 2017). Most landscape characteristics such as rivers, water systems and utilities have much smaller dimensions than 100-500 kilometers. Climate patterns, such as convective clouds and coastal breezes, could be captured after appropriate downscaling.

Downscaling requires a variety of approaches, each with its benefits and drawbacks. Several methods are available for downscaling meteorological information. They can be broadly classified as dynamical and statistical downscaling focusing on regional climate simulations[(Kaur et al. 2020)(Nobre et al. 2001; Díez et al. 2005; Shukla and Lettenmaier 2013; Xue et al. 2014)]. The statistical downscaling methods [(von Storch et al. 1993; Zorita and von Storch 1999; Wilby and Dawson 2013; Sahai et al. 2017)] have been applied in several studies mainly focussing on the downscaling of precipitation [(von Storch et al. 1993; Vrac and Naveau 2007; Benestad 2010; Sahai et al. 2017)].

Dynamical downscaling involves using the initial and boundary conditions from a global model and then running a high-resolution regional model to generate a local forecast. For example, city-level forecasts, which are often computationally expensive and time-consuming, require high-performance comput resources [ (Benestad and Haugen 2007; Benestad 2010; Kaur et al. 2020)]. On the other hand, statistical downscaling uses the outputs from the global dynamical models or observations as inputs to statistical models (ranging from simple univariate to complex multivariate schemes) to generate high resolution (e.g., city-scale) information.

Statistical downscaling of rainfall (observation or forecast) is a popular (computationally) low-cost method (so-called the poor man's approach) to improve the rainfall information on regional scales and is made available to stakeholders and decision-makers[(Salvi et al. 2013; Vandal et al. 2017) ]. The world has moved towards Artificial Intelligence (AI) and Machine Learning (ML) techniques for statistical downscaling. Several studies show that AI-based downscaling methods could be suitable for rainfall downscaling, which are often very non-linear and include multi-scale stochastic noise. Tripathi2006 has shown the strength of the Support Vector Machine (SVM) method in the downscaling of precipitation data. The recent developments in Single Image Super-Resolution (SR) using Deep Learning [(Dong et al. 2015)] offers promising hope in downscaling weather model outputs to high resolution. Figure 1c depicts a visual representation of the downscaling principle. This paper introduces a deep learning-based method for precipitation downscaling to generate high-resolution precipitation data over the Indian region during the summer monsoon (June-September) season, which can provide local projections. The next section of this paper describes methods followed by the results obtained in section 3. Conclusion and discussion are provided in section 4.

## 2. Methods

This study uses Super-resolution based Convolutional Neural Networks (CNN), known as SRCNN, following[(Vandal et al. 2017, 2019)] wherein a low-resolution input image is used to map to a high-resolution image accurately. SRCNN allows end-to-end mapping to be optimised and is faster than other super-resolution (SR) methods. It uses a low-resolution image in the first step, upgraded to the higher resolution size by using bicubic interpolation. Suppose the interpolated image is referred to as Y, SRCNNs task is to retrieve from Y an image F(Y) which is close to the high-resolution ground truth image X. We still call Y as a" low resolution" image as the pixel level information is inaccurate in Y relative to X. (Dong et al. 2015) showed that CNNs could accurately extract patches, conduct non-linear mapping, and reconstruct a low-resolution input to a high-resolution image.

Traditional SR approaches are used to boost the resolution of a factor of 2-4; statistical downscaling demands a conservative resolution increase of 8-12 factors. Similar to (Vandal et al. 2019), we followed an alternative approach where instead of explicitly increasing the resolution to 8-12x, it is achieved in many steps and known as stacked SRCNN.

We used stacked SRCNNs to boost the resolution to such a high degree that each SRCNN improves its resolution by one factor. The model can learn spatial pairs on multiple levels through this approach, requiring less complexity in the spatial representations. The ability to arbitrarily improve image resolutions from the lower resolution enables input/output pairs to train stacked SRCNNs on several scales. However, since the SRCNN output is again the SRCNN input, an unwanted network error may be distributed over the stacked layers when training a case by case model.

In general, one only requires a low resolution (LR) image to approximate a high resolution (HR) image while using an SR on pictures. Nonetheless, during statistical downscaling (SD), high-resolution data may be used to estimate the HR images, which coincide with this LR image. For example, we can use two types of inputs for precipitation downscaling, such as maps of LR rainfall and static topographical features. Since topographical characteristics are defined in very high resolution and do not usually change over time, they can be used for any scaling factor. The method which uses topographies as an additional input to stacked-SRCNN has been called the `DeepSD' by (Vandal et al. 2019). A sketch in Figure 2 provides an overview of SRCNN and DeepSD methods. Table 1 provides the data training parameters used in this work.

## 3. Data

In this work, we have primarily used rainfall data obtained from several sources. The theory is first developed using the daily IMD (Indian Meteorological Department) gridded data. Data on two spatial resolutions are used. The low resolution $1^o X 1^o$ spatial resolution data is used as input to the AI/ML network, and the $0.25^o$ X $0.25^o$ is used for verification (ground truth) data. The low resolution ($1^o X 1^o$) data is generated from the high-resolution data using linear interpolation. The IMD data is prepared from the Indian land station data described in [(Rajeevan et al. 2006, 2008; Pai et al. 2014)]. For each day, The $1^o$ data has (*latitude X longitude*) dimension of (33x35), and $0.25^o$ data has the (*latitude X longitude*) of dimension (129x135). Apart from IMD daily gridded data, the TRMM data (Huffman et al. 2007) of resolutions ($0.25^0 X 0.25^0$) was also used to test these algorithms. We have also used the *Fifth generation of ECMWF atmospheric reanalysis of the global climate* (ERA5) rainfall data (Hersbach et al. 2020). This data was downloaded from Copernicus Climate Data Server (CDS) to demonstrate the potential application of the AI/ML based method for the postprocessing (i.e., downscaling and bias-correction) of reanalysis rainfall data as a potential application of the concept developed using IMD data over the Indian region.

## 3.1. Data preprocessing

Original data obtained from IMD has values only on the land surface area. However, for applying the CNN algorithm, square/rectangular shape data is required for 2D convolution. The 2D convolution kernels are generally square and are applied to the whole matrix. The original data had a triangular shape, based on the data availability over the Indian region signifying no values on the ocean area or outside the political boundaries of Indian landmass were filled with significantly large negative amounts in the IMD gridded dataset. First of all, negative values outside Indian region limits were converted into 'NaN'. Then additional 'NaN' values were padded to make matrix size 140x140 from 129x135 originally, as shown in fig.4. A similar transformation is used to make the square shape for TRMM data and ERA5 rainfall data.

## 3.2 Training and testing

Three methods described in the previous section (Sec.2) were employed on IMD data to develop a model for creating high-resolution data. The training was done for 30 years on $1^o$ x$1^o$ IMD data, i.e., from the year 1975-2004 for the June to September period. A systematic analysis was carried out to select the best data training architecture for each model. The developed models were tested on five years of monsoon season data (122 days of June-September) from 2005-2009. The resulted outputs were compared with ground truth, i.e., $0.25^o$ IMD gridded data. Similarly, the TRMM and ERA rainfall data is also partitioned as training and test period data.

## 3.3 Verification metrices

In this study, we use two types of verification metrices. The first metric is the traditional mean square error (MSE) and pattern correlation coefficients (cc) based approach and the second metric is based on the multivariate statistical analysis approach. The fidelity of the AI/ML model output data is first tested for the Indian region by plotting the probability density function (*pdf*) of ***cc*** and ***MSE*** computed for all the grid points where data is available or undefined. The ***cc*** and ***MSE*** for each grid point are computed based on the 5 years X 122 days =610 days of data. PDFs of ***cc*** and ***MSE*** are plotted for the grid points where the metric is computable (i.e., a grid point out of the total 129X135=17415 grid points which do not have undefined or NaN attribute for all the 610 days). In the next step, we have calculated the principal components (***pc***) and the Eigenvectors of the covariance matrix (i.e., empirical orthogonal functions, aka ***EOF***, (Hersbach et al. 2020) for the spatial data over the Indian region. The first three *pcs* and *EOFs* are plotted to see that the principal components' dominant spatial variance and the temporal evolutions are faithfully represented by outputs from the AI/ML-based networks.

## 4. Results and Discussion

The models developed have been employed to two different sets of data, as mention in section 2. Here, we provide details of the analysis done on these data.

## 4.1 Downscaling of IMD data

For visual comparison, 2D images of rainfall amount on a particular day in a year were compared, as shown in figure 5 These days are representative of heavy precipitation on the east coast and west coast monsoon rainfall.

In figure 5, the rainfall amount obtained from three methods, viz., SRCNN, Stacked-SRCNN, and DeepSD, are compared with ground truth and traditional linear interpolation. Panel (b) is the original input having resolution $1°$, and panels (a) and (c) represents linear interpolation to $0.25°$ and ground truth of $0.25°$ resolution, respectively. Outputs obtained from three methods are represented by panels (d), (e) and (f), which are to be compared with panel (a) and (c). Rectangles indicate the prime region of maximum rain in the southwestern part of India. The results obtained from all three methods captured better rainfall than the traditional linear interpolation method (panel a). This comparison of findings suggests that DeepSD is a better tool for downscaling.

Similar comparisons were also made for extreme case years that is for drought year and extra monsoon year. In those comparisons (not shown here) too, DeepSD was found to the best method. However, it does not sound right to draw generalized conclusions based on these few comparisons. Therefore, we have carried out statistical analysis and compared two measures: pattern correlation (**CC**) and mean square error (**MSE in mm$^2$/day**) for the three methods with *pdf* plotted in **Figure 4** as discussed in sec.3.3.

In this comparison also, the deepSD model was found to be better among the three methods as it shows the highest frequency (probability density) of maximum correlation value and the lowest MSE peak frequency, as seen in the **Figure. 4**. Interestingly, using this metric, we find that all these methods have performed better than conventional linear interpolation (blue curve).

Furthermore, as another test, the entire data was divided into four quartiles based on the count of grid points where rainfall values are more than 30 mm. Model results were compared in the upper quartile also ensure that they are consistent. **Figure 5** contains comparisons for PDFs of pattern correlation as well as MSE. Similar to results for entire data, here also deepSD stood as the best method among all three.

In order to see if the AI/ML based downscaled data has preserved the spatial coherence and the dominant mode of variability is represented adequately, we plotted the first three EOFs of the downscaled data together with ground truth, which are depicted in **Figure 6.** It can be seen that the dominant modes of all the methods represent the dominant spatial covariance pattern for the observation, although the first three modes overestimated the spatial variance (mentioned at the top tight of each panel). The **r$^2$** (square of correlation) values of the pcs are shown in **Table 1a.**

| AI model name | Mode 1 | Mode 2 | Mode 3 |
|---|---|---|---|
| (DeepSD, IMD) | 92 | 88 | 59 |
| (Stack-SRCNN, IMD) | 92 | 88 | 67 |
| (SRCNN, IMD) | 92 | 88 | 65 |

**Table-1a:** Square of Correlation coefficient (r$^2$,%) of AI/ML models with IMD ground truth PC.

It may be seen that up to mode 2, all the models capture the temporal variability (represented by the CC of the PCs). However, as we go to the higher modes, the temporal variability is not faithfully represented. One of the drawbacks of EOF based modes is that higher modes may lack physical relevance with the modes are more geometrical in nature and hence MSE and pattern CC based PDF would be useful to represent the variability of total fields. Thus based on these two metrices, we can infer that the downscaled AI/ML models captured the large-scale patterns, and the model preserves the temporal variability of the ground truth data.

## 4.2 Downscaling of TRMM Data

The application of the above three methods on IMD data assured that 'DeepSD' is the best method for downscaling. This method was applied to TRMM data to downscale from 0.5o to 0.25o resolution to re-confirm this analysis.

The total data used in the case was 17 years, i.e., 1998-2015. Out of which 15 years of the data were used for model training and the rest two years (2014- 2015) data for model validation. Details of parameters used for training and testing this data are the same as those in Table 1. Results were then compared with traditional linear interpolation.

The correlation and MSE pdfs were computed and compared with linear interpolation. It was found that DeepSD has a better correlation and less MSE than linear interpolation shown in **Figure 7**.

## 4.3 Application for Model data postprocessing

### 4.3.1. Model data postprocessing:

We have applied this model for the postprocessing of ERA 5 reanalysis data. In this case, the model's training was done using ERA5 data and IMD gridded data as ground truth. The training data was 25 years, in particular, from 1979-2004. The testing was done on the JJAS period from 2005-2009. Similar to IMD and TRMM data, in this case, deepSD found to be better than SRCNN. A comparison of correlation and MSE using original ERA5 data, SRCNN, and deepSD is shown in **Figure 8**.

Similar to the IMD rainfall postprocessing, we also use the EOF-based metric to evaluate the dominant mode of variability's spatial patterns. **Figure 9** shows that compared to the raw ERA5 rainfall, the variance distribution of ERA5 rainfall post-processed using AI/ML methods is better. Especially, the variance over the northern part of India, northwest India, and Kasmir, captured in a better way after AI/ML-based postprocessing. Also, the **$r^2$** (square of the correlation coefficient) improves a lot (see Table-1b) in the AI/ML-based post-processed data compared to the raw ERA5 data.

**Table 1b**: Square of Correlation coefficient ($r^2$,%) between AI/ML-based ERA5 post-processed rainfall and IMD rainfall

|  | Mode 1 | Mode 2 | Mode 3 |
|---|---|---|---|
| (Raw-ERA5,IMD) | 58 | 43 | 18 |
| (DeepSD-ERA5, IMD) | 67 | 68 | 18 |
| (SRCNN-ER5,OBS) | 68 | 59 | 14 |

### 4.3.2. Scale-free downscaling:

These results proved that DeepSD is the best method among the three methods discussed in this work. Based on our experience in 1x1 to 0.25x0.25 degree downscaling, we hypothesise that SRCNN or DeepSD based method will provide the realistic amplitude and intensity of extreme events without any loss of information. We applied this model to generate 0.125° data using 0.25° data obtained from IMD observation.

A comparison of the resulted high-resolution map obtained from traditional linear interpolation and our models, namely, SRCNN and DeepSD, are shown in **Figure 10**.

## 5. Conclusion

This study has employed an application based on AI/ML-based techniques, namely, SRCNN & DeepSD, to downscale the Indian region's rainfall information. The analysis shows, DeepSD based method has a better correlation and lowest MSE than the conventional linear interpolation approaches. The AI/ML-based approach provides better information on rainfall intensity as compared to linear interpolations. It is well known that the AI/ML-based method has the advantage of superior gradient or edge detection than linear techniques.

The downscaling of precipitation based on this hypothesis of superior edge detection by AI/ML-based methods could improve the spatial distribution of downscaled rainfall. This hypothesis has been verified for rainfall data, and the benefits of AI/ML-based downscaling has been demonstrated.

It is shown that the spatial discontinuity in amplitude and intensity rainfall patterns is the main obstacle in the downscaling of precipitation over the Indian region. The AI/ML method applied to the station derived meteorological data (IMD) has the advantage of preserving this spatial discontinuity of rainfall data without any loss in intensity. The model has been used on two different data types; namely, IMD gridded data obtained from the ground (station) observation and the satellite-derived TRMM data. In both cases, the best correlations were found with 'DeepSD,' and the lowest MSE was obtained using this method.

Similarly, it is found that the AI/ML-based model can be used in a proficient manner to bias correct the reanalysis rainfall. The post processed ERA5 rainfall data using the AI/ML techniques show much better distribution of spatial covariance and the temporal variance when compared with observation. Thus, the AI/ML-based postprocessing and downscaling is proposed to be an efficient and skilful method for rainfall data over the Indian region.

After validating these methods, we have generated very high resolution ($0.125^0$ x $0.125^0$) rainfall maps over the Indian region using input data of ($0.25^0$ x $0.25^0$) resolution. This work can be the first step towards developing deep learning-based meteorological data downscaling models.

**Acknowledgment**
The authors would like to thank Dr. David John Gagne, NCAR, for a constructive discussion and suggestions on this manuscript.
This work was done on HPC facility provided by the Ministry of Earth Sciences at IITM Pune. The details of HPC can be found on prtyush.tropmet.res.in. IMD rainfall data is obtained from India Meteorological Department, Pune. The TRMM data is available at https://disc.gsfc.nasa.gov/datasets/TRMM_3B42_Daily_7/summary. The ERA5 rainfall data is downloaded from https://cds.climate.copernicus.eu/

**Conflict of interest**
The authors declare that they have no conflict of interest.

**Funding:** This work is a part of a student project completed at IITM. The computational facility was provided by IITM. No additional funding support was received.

**Author Contribution:**

**Concept and design**: B.K , R.C. and M.S conceptualize the idea of problem statement and finalized the algorithms for methods used in this study.

Drafting of manuscript: B.K. and R.C. mainly contributed in manuscript writing. M.S. also helped to finalize the text for manuscript.

**Code development:** N.C. and K.D developed code. M.S. and A.B. supervised the code development and plotting of figures.

**Acquisition of Data:** R.C. and M.S. contributed in data collection and pre-processing.

**Critical revision**: R.C and M.S did critical revision of the manuscript.

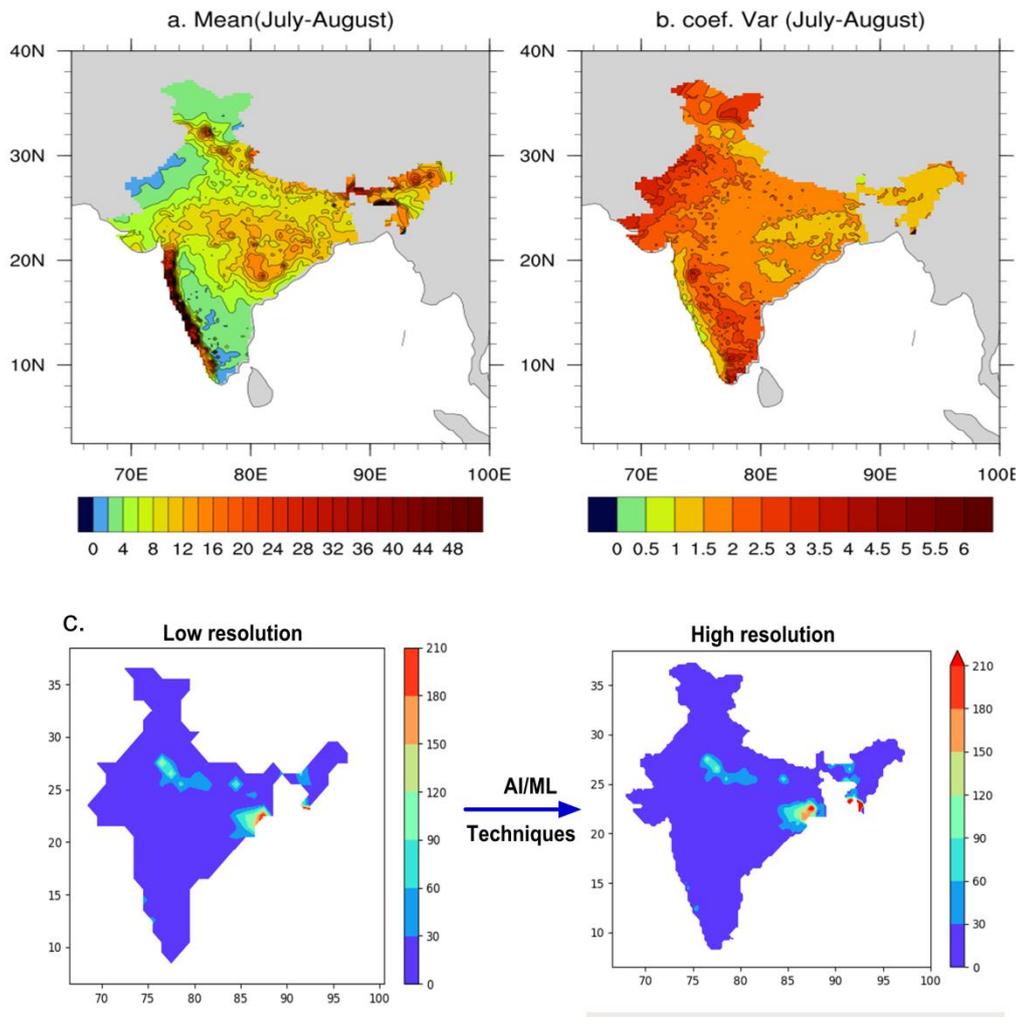

**Figure 1**: Upper panel plots showing the July-August mean rainfall millimetres per day (a), and coefficient of Variation (CoV) (b) for the years 1981-2010. The region with low mean has high dispersion and high mean has low dispersion. There are sharp spatial gradients of CoV long the orographic variation zones, especially in the Western Ghats and Northern Himalayan belts. Lower panel plot (c) illustrate concept of downscaling of rainfall data. This units of rainfall in legends are millimetre/day.

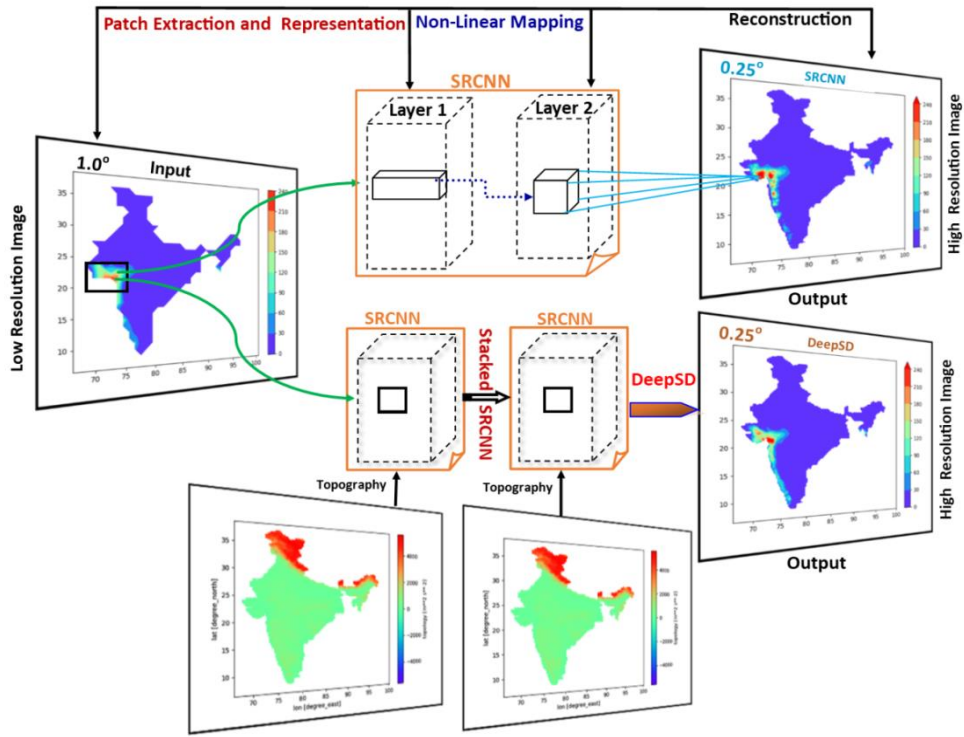

**Fig 2**: An overview of SRCNN and DeepSD methods. In the DeepSD, the downscaling is done in steps rather than a direct 4x or 8x resolution. Also, DeepSD used multivariable inputs.

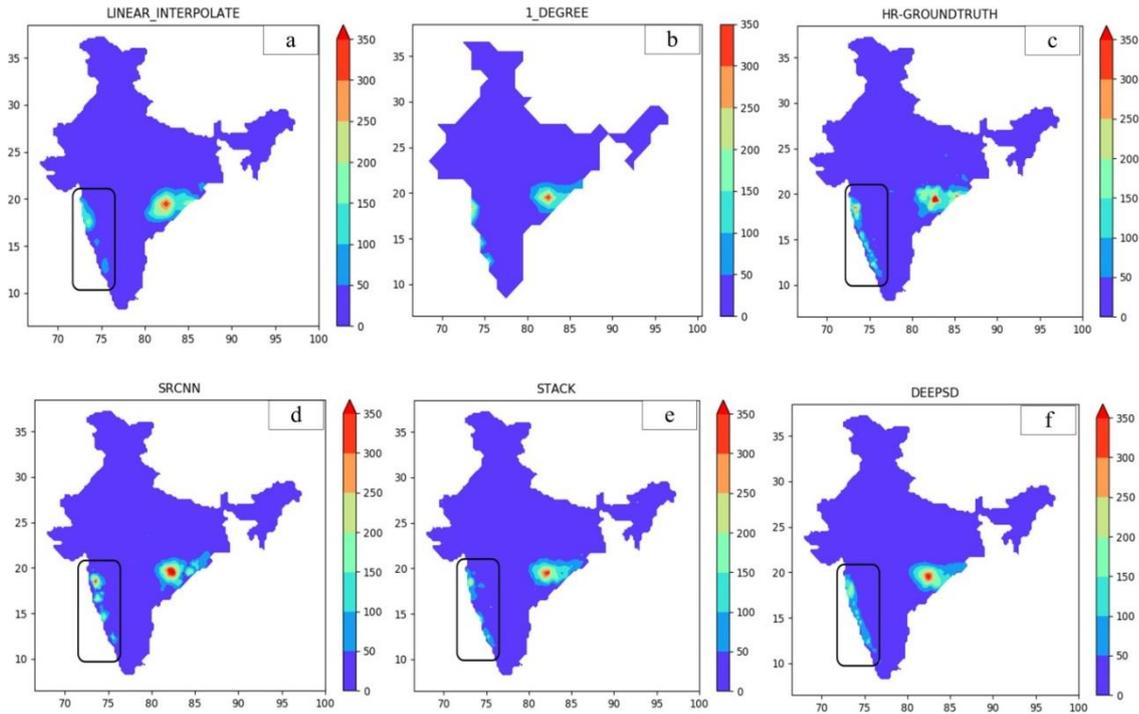

**Fig 3:** Comparison of downscaling output resulted from three AI models applied in this study with traditional linear interpolation. The data is for the year 2006 and day 183. The rainfall amounts shown are in millimeters (mm) per day.

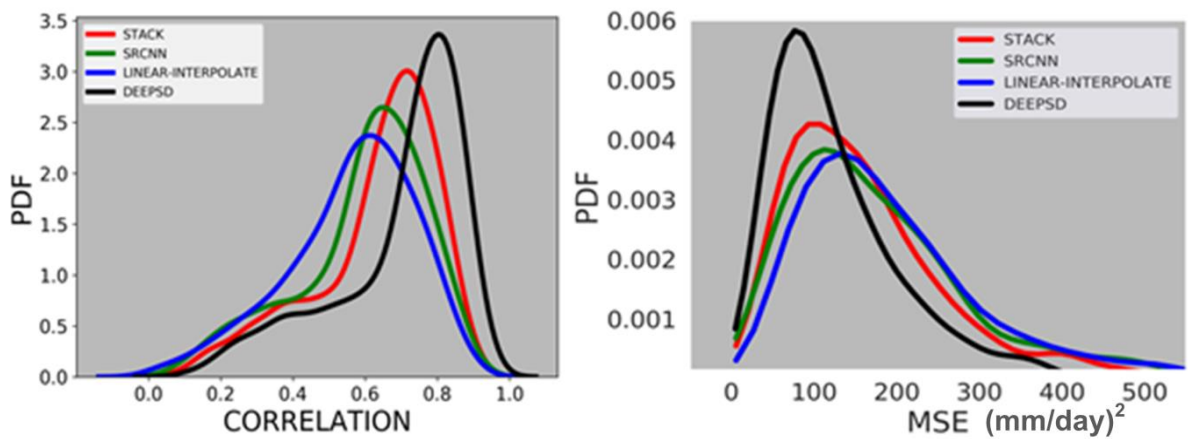

**Fig 4:** Comparisons of correlation PDF and MSE $(mm/day)^2$ using three methods applied on IMD data.

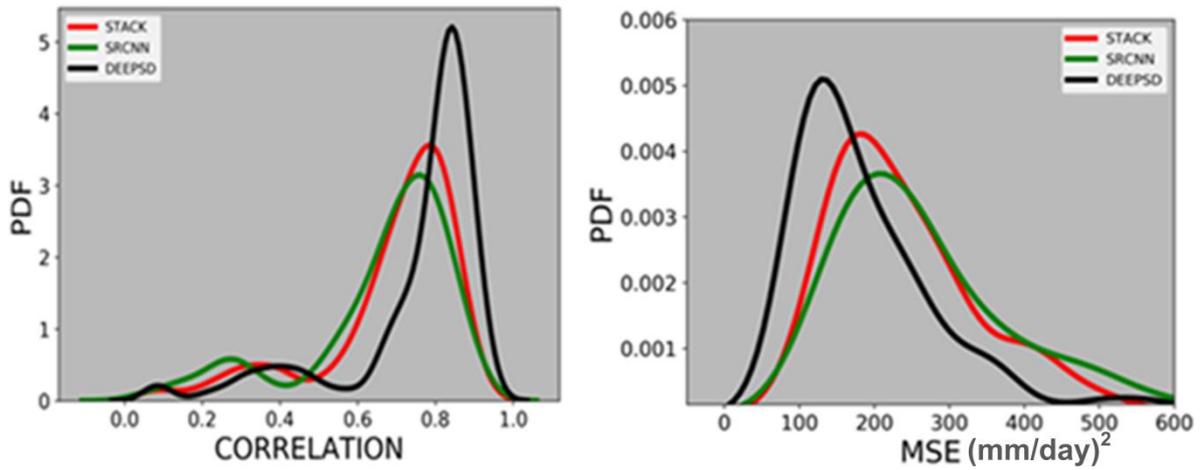

**Fig. 5** Comparisons of correlation PDF and MSE $(mm/day)^2$ using three methods on upper quartile only.

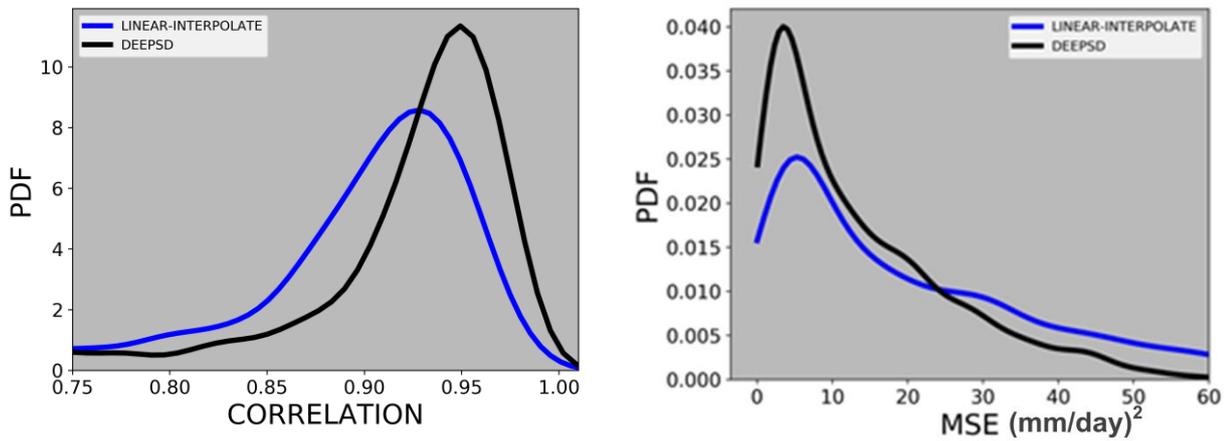

**Fig. 6** Comparison of correlation and MSE $(mm/day)^2$ pdf for TRMM Data. Correlation calculated from DeepSD method is better than linear interpolation. MSE is slightly less in the case of DeepSD than linear interpolation.

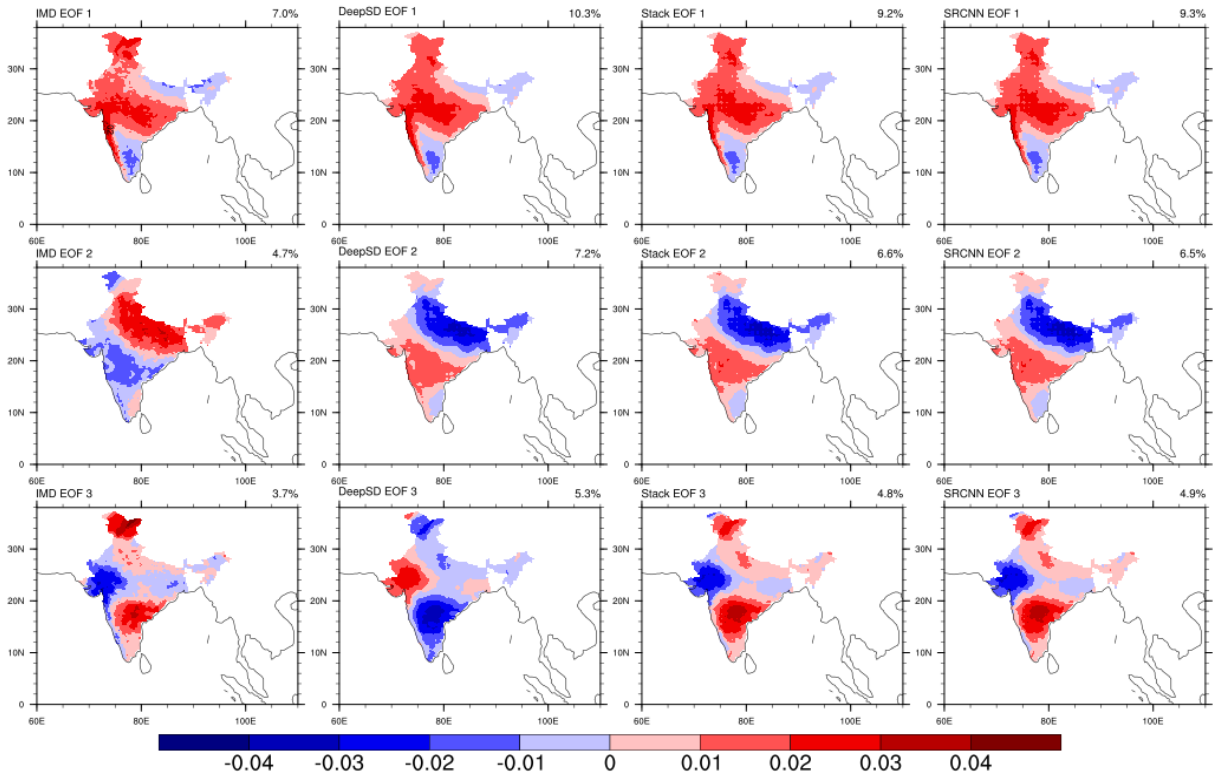

**Fig7**: First three empirical orthogonal function (EOF) of IMD ground truth data (1st-row panels) and AI/ML based models as mention at the top of each model (2nd row is for DeepS, 3rd for Stacked SRCNN, and last depicts SRCNN EoFs). The variance explained by each mode is also mentioned at the top-right of each panel. The amplitudes of shading are arbitrary units.

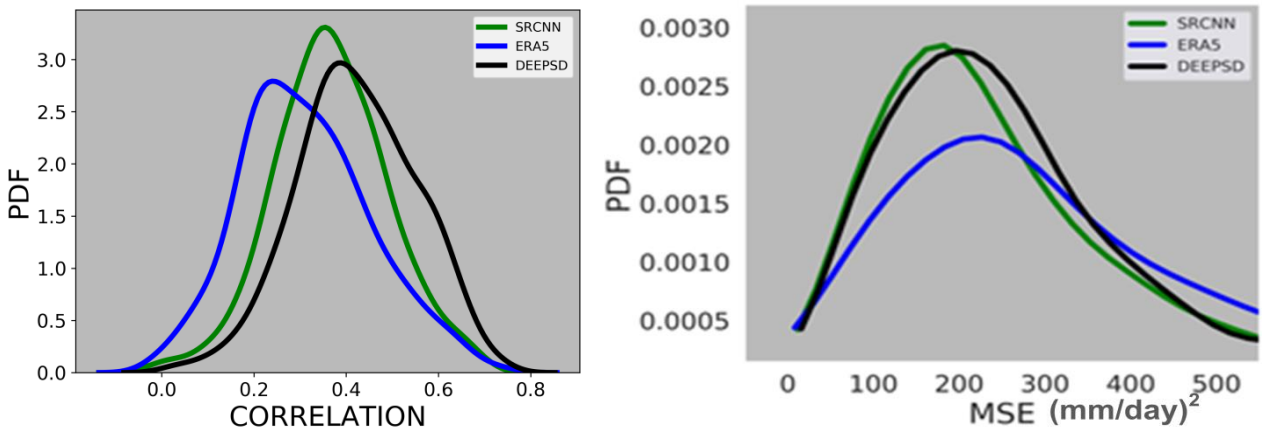

**Fig 8:** Comparison of correlation and MSE (mm/day)$^2$ pdf for TRMM Data. The blue color line is for original ERA5 data.

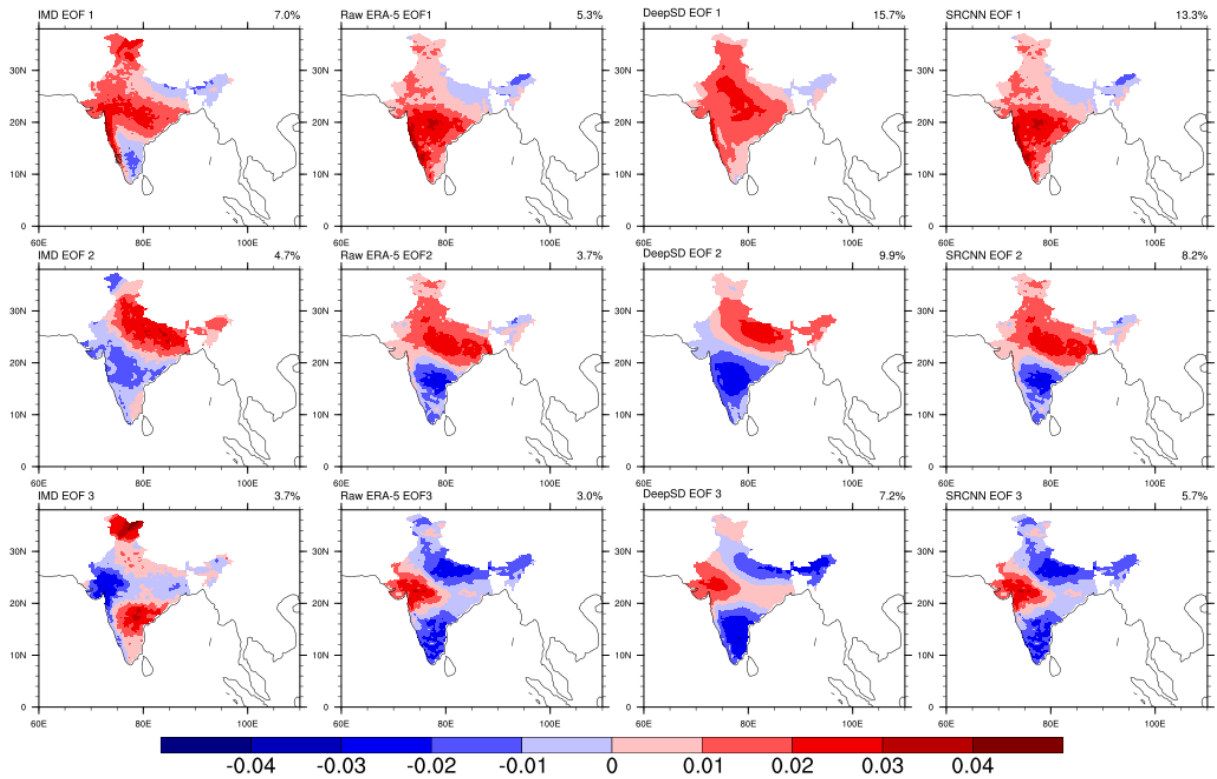

**Figure 9**: The first row (3 panels from top to bottom) represents the EOFs for IMD ground truth; 2nd row shows the raw ERA-5 reanalysis rainfall data. The EoFs for AI/ML based post-processed data using the ERA-5 reanalysis data as input (mentioned at the top of each panel) are shown in 3rd and 4th rows. The variance explained by each of the modes is also mentioned at the top-right of each panel. The amplitudes of shading are arbitrary units.

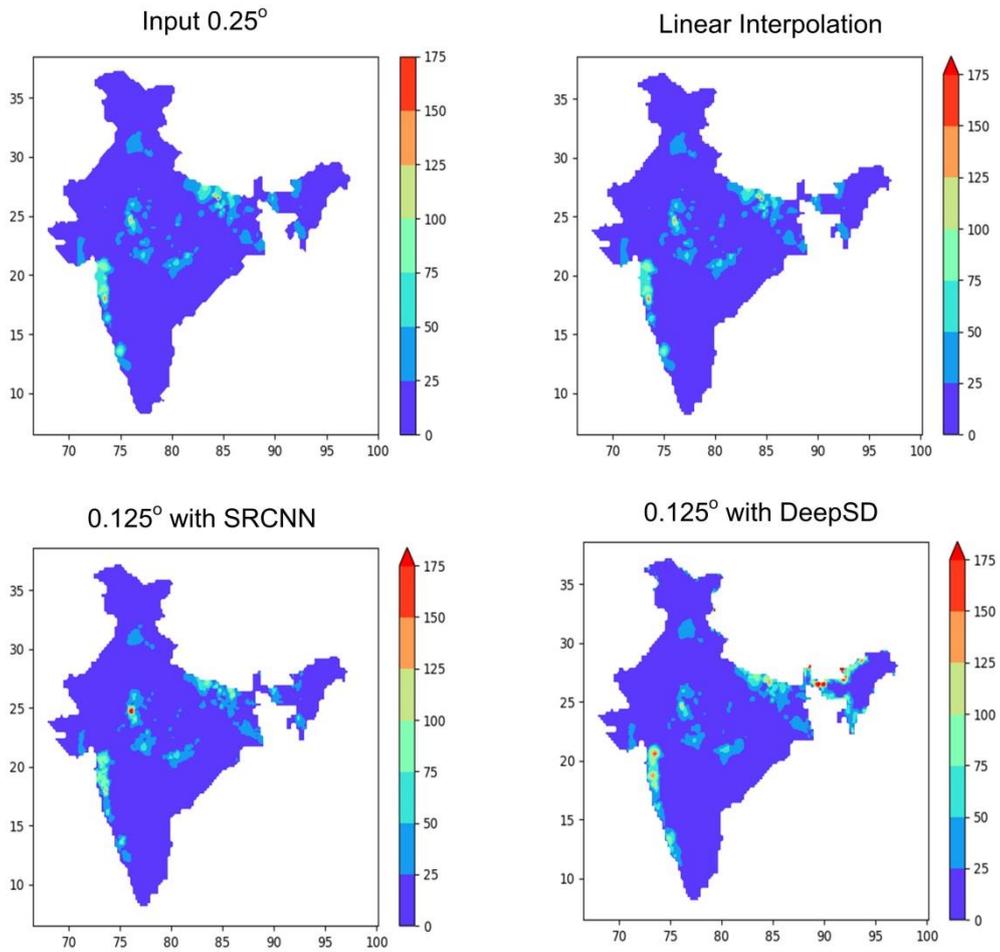

**Fig.10** Application of developed models. A high resolution (0:125o _0:125o) rainfall map over Indian region (landmass) was generated using SRCNN and DeepSD methods. The map resulted from the DeepSD shows better rainfall intensity. Since no such high-resolution observation data is available at IMD, it cannot be validated with observation.